\documentclass[11pt]{article}
\pdfoutput=1
\usepackage{amssymb,amsmath,mathrsfs,enumerate}
\usepackage{graphicx,rotate,multicol}
\usepackage{caption,verbatim}
\usepackage{cite}
\usepackage[colorlinks=true,
linkcolor=red,
urlcolor=blue,
citecolor=blue]{hyperref}
\usepackage{color}
\usepackage{textcomp} 
\usepackage{setspace}
\usepackage{subfig}

\evensidemargin=\oddsidemargin
\def\mET{E_T \hspace{-1.2em}/\;\:}

\allowdisplaybreaks

 \usepackage[normalem]{ulem}
 
 \definecolor{darkgreen}{cmyk}{1,0,1,0.4}
 \def\hldg#1{\textcolor{darkgreen}{\large\textsl{#1}}}

\def\t#1#2{$#1 \times 10^{#2}$}

\begin{document}

\begin{flushright}
{\small HRI-RECAPP-2016-003}  
\end{flushright}

\begin{center}
{\Large \bf Little Higgs after the little one} \\
\vspace*{1cm} {\sf Debajyoti Choudhury$^{a,}$\footnote{debajyoti.choudhury@gmail.com}, 
	~Dilip Kumar Ghosh$^{b,}$\footnote{tpdkg@iacs.res.in},
        ~Santosh Kumar Rai$^{c,}$\footnote{skrai@hri.res.in},
        ~Ipsita Saha$^{b,}$\footnote{tpis@iacs.res.in}} \\
\vspace{10pt} {\small } {\em $^a$ Department of Physics and Astrophysics,  
	University of Delhi, Delhi 110007, India \\
	$^b$Department of Theoretical Physics, 
	Indian Association for the Cultivation of Science,\\
	2A $\&$ 2B, Raja S.C. Mullick Road, Kolkata 700032, India \\
	$^c$Regional Centre for Accelerator-based Particle Physics, \\
	Harish-Chandra Research Institute, Chhatnag Road, Jhusi, Allahabad 211019, India}  
\end{center}
\normalsize

\begin{abstract}
At the LHC, the Littlest Higgs Model with $T$-parity is 
characterised by various production channels. If the $T$-odd quarks 
are heavier than the exotic partners of the $W$ and the $Z$, then 
associated production can be as important as the pair-production 
of the former. Studying both, we look for final states comprising at least 
one lepton, jets and missing transverse energy.
We consider all the SM processes that could conspire to contribute 
as background to our signals, and perform a full detector level 
simulation of the signal and background to estimate the 
discovery potential at the current run as well as at the scheduled
upgrade of the LHC. We also show that, for one of the channels,
the reconstruction of two tagged $b$-jets at the Higgs mass 
$(M_h = 125~{\rm GeV})$ provides us with an unambiguous hint for 
this model.
\end{abstract}

\bigskip


\section{Introduction}\label{intro} 
The Standard Model (SM) of particle physics provides an admissible
explanation for the electroweak symmetry breaking (EWSB) mechanism
that seems to be in accordance with all observations till
date including the electroweak precision tests.  The discovery
of the long sought Higgs boson at the Large Hadron Collider
(LHC)~\cite{Aad:2012tfa,Chatrchyan:2012ufa} completes the search for
its particle content, and the current level of
agreement of this particle's couplings to the other SM particles is a
strong argument in favour of the model. In spite of such a triumph,
the SM is beset with unanswerable problems, whose resolution requires
the introduction of physics beyond the domain of the SM. One such issue
pertains to the smallness of the Higgs mass, which is unexpected as
there exists no symmetry within the SM that would protect the Higgs
mass from radiative corrections. This extremely fine-tuned nature of
the SM is termed as the {\em Naturalness Problem} and many scenarios
beyond the SM (BSM) such as supersymmetric theories, extra dimensional
models and little Higgs models have been proposed as solutions.

In the little Higgs models, the Higgs boson is realized as a pseudo
Goldstone boson of a new global symmetry
group~\cite{ArkaniHamed:2001nc,ArkaniHamed:2002pa,ArkaniHamed:2002qx}.
With the Higgs mass now being proportional to the extent of the {\em
  soft breaking} of this symmetry, the relative lightness can,
presumably, be protected. The minimal extension of the SM based on
the idea of little Higgs scenario is the Littlest Higgs
model~\cite{ArkaniHamed:2002qy,Schmaltz:2002wx}, which is essentially
a non-linear sigma model with a global $SU(5)$ symmetry that breaks
down to $SO(5)$ at some scale $\Lambda$ on account of a scalar field
vacuum expectation value $f\approx\Lambda/4\pi$.  A subgroup of the
$SU(5)$, namely $[SU(2)\times U(1)]^2$, is gauged, and the
breaking mechanism is such that the local symmetry
spontaneously breaks into its diagonal subgroup which is identified
with the SM gauge group $SU(2)_L\times U(1)_Y$.

Unlike in supersymmetric theories, the cancellation of the leading
correction to the Higgs mass square occurs here between contributions
from particles of the same spin\footnote{Note that the subleading
contributions do not cancel.}. For example, the $W/Z$ contributions
are cancelled by those accruing from the extra gauge
bosons. Similarly, it is the exotic partner of the top quark that is
responsible for cancelling the latter's contribution.  The collective
symmetry breaking mechanism ensures that no quadratic divergence
enters in the Higgs mass before two loops.  Although, technically, the
little Higgs models, unlike supersymmetry, are not natural (for the
stabilization of the scale $\Lambda$ is not guaranteed and has to be
ensured by other means), the inescapability of this extra loop
suppression ameliorates the fine tuning to a great degree rendering it
almost acceptable.

On the other hand, the very presence of these extra particles results
in additional contributions to the electroweak precision observables \cite{Csaki:2002qg,Hewett:2002px,Csaki:2003si,Chen:2003fm,Kilian:2003xt,Han:2004az,Marandella:2005wd}, 
and consistency 
with the same requires that the scale $f$ should be above a few TeVs, thereby introducing the
`little hierarchy problem'.  These constraints can, however, be
largely avoided with the introduction of a new discrete symmetry,
namely `$T$-parity', under which all the SM particles are even while
all the new particles are odd.  This forbids the mixing between the SM
gauge bosons and the heavy $T$-odd gauge bosons at the
tree-level, thereby preserving the tree-level value of the electroweak
$\rho$-parameter at unity~\cite{Agashe:2014kda}. 
The Littlest Higgs model with $T$-parity
(LHT)~\cite{Cheng:2003ju,Cheng:2004yc,Low:2004xc,Hubisz:2004ft,Hubisz:2005tx},
thus, solves the little hierarchy problem and has the additional
advantage that the lightest $T$-odd particle (which naturally happens
to be electrically neutral and color-singlet) can be a good cold Dark
Matter (DM) candidate~\cite{Birkedal:2006fz,Asano:2006nr,Chen:2014wua}. 

The LHT model, like any other BSM scenarios also has interesting
phenomenological implications with its own set of non standard
particles.  In light of the Higgs discovery a detailed analysis of the
model has been considered at run-I of the LHC~\cite{Reuter:2013iya}.
In this work, we investigate a few of the most likely signatures of
LHT that could be observed at the current run of LHC with
$\sqrt{s}=13$ TeV as well as predictions for the possible upgrade to
$\sqrt{s}=14$ TeV.  As the discrete $T$ symmetry forbids single
production of any of the $T$-odd particles, they must be pair produced
at the LHC.  While the pair production of $T$-odd gauge boson
$(W_h^\pm)$ has been studied in
Refs.\cite{Cao:2007pv,SongMing:2012gb,Wen:2013xb,Liang-Wen:2014fla,Cao:2015cdb}, 
unless the Yukawa couplings are very large, the production rates are expected to be
higher for processes involving the exotic quarks. Here, we consider
the signals generated from the associated production of heavy $T$-odd
quarks with heavy $T$-odd gauge bosons. We eschew the simplistic
possibility that the exotic quark decays directly into its SM
counterpart and the invisible $A_h$ (relevant only for a limited part
of the parameter space and considered in Ref.\cite{Choudhury:2012xc})
and consider (the more prevalent and more complicated) cascade decays
instead.  We concentrate on final states---for LHC run II---comprising
leptons and jets accompanied by large missing transverse energy,
while noting that the pair-production of the $T$-odd quarks also
contributes significantly owing to their larger cross-sections. 
For a large part of the allowed parameter space, the $Z_h$ boson
dominantly decays to a Higgs boson and $A_h$. This, consequently,
gives rise to two $b$-tagged jets, thereby proffering the interesting
possibility of reconstructing the Higgs mass and validating the decay
chain predicted by the model.  Performing a detailed collider analysis
while taking all the relevant SM backgrounds into consideration, we
explore the possibility of probing the model parameter space at the
current run of the LHC.

The paper is organized in the following manner: In
Section~\ref{model}, we begin with a brief description of the LHT
model. In Section~\ref{analysis}, we describe our analysis strategy
and explore the discovery possibilities at the LHC run II. Finally, in
Section~\ref{conclusion}, we summarize our findings and conclude.

\section{The Littlest Higgs model with T-parity}\label{model} 
Consider a non-linear sigma model with a global $SU(5)$ symmetry of
which the subgroup $G_1 \otimes G_2$, with each $G_i \equiv SU(2)_i
\otimes U(1)_i, (i =1, 2) $, is gauged. If $\Sigma$ is a dimensionless
scalar field transforming under the adjoint representation, its
kinetic term could be parametrized as
\begin{equation}
{\cal L} = \frac{f^2}{8} \rm Tr(D_\mu \Sigma)^\dagger (D^\mu \Sigma)\,, \label{lag}
\end{equation}
where the covariant derivative $D_\mu$ is defined through
\begin{equation}
D_\mu \Sigma = \partial_\mu \Sigma - i \sum_{j=1}^2 [g_j W_j^a(Q_j^a \Sigma + \Sigma Q_j^{aT}) + g_j^\prime B_j (Y_j\Sigma +\Sigma Y_j)] \ . 
\label{kin}
\end{equation}
The gauged generators can be represented in the convenient form 
\begin{equation}
\begin{array}{rcl c rcl}
Q_1^a &=& \displaystyle \frac{1}{2} \, 
     \left(\begin{array}{ccc}
          \sigma^a & 0 & 0  \\
           0 & 0 & 0  \\
           0 & 0 & 0 
     \end{array}\right)
& \qquad \quad &
 Y_1 & = & \rm diag(3,3,-2,-2,-2)/10 \\[5ex]
Q_2^a &=& \displaystyle
    \frac{1}{2} \, \left(\begin{array}{ccc}
                          0 & 0 & 0  \\
                          0 & 0 & 0  \\
                          0 & 0 & -\sigma^{a*}
                          \end{array}\right)
& & Y_2 & = & \rm diag(2,2,2,-3,-3)/10 \ , 
\end{array} 
\label{generator}
\end{equation}
where $\sigma^a$ are the Pauli matrices. The imposition of a $Z_2$
symmetry ($T$-parity) exchanging $G_1 \longleftrightarrow G_2$ (and,
naturally, the corresponding quantum numbers for all the fields in the
theory), requires that
\begin{equation}
g_1 = g_2 = \sqrt{2} \, g  \ , \qquad
g_1^\prime = g_2^\prime = \sqrt{2} \, g^\prime \ ,
\end{equation}
where $g$ and $g'$ would shortly be identified with the SM gauge
couplings.
The 
global $SU(5)$ symmetry is spontaneously 
broken down to $SO(5)$ by
the vacuum expectation value (vev) $(\Sigma_0)$ of the scalar field
$\Sigma$ at the scale $f$, viz.
\begin{equation}
\Sigma_0 = \left(\begin{array}{ccc} 
& & \mathbf{I}_{2\times 2}  \\
& 1 &  \\
\mathbf{I}_{2\times 2} & &
\end{array} \right) \ , 
\label{sig0}
\end{equation}
thereby leading to 14 Goldstone bosons.  The field $\Sigma$ can be
expanded around the vev as
\begin{equation}
\Sigma(x)= e^{2i\Pi/f} \Sigma_0 \,, \label{nlsm}
\end{equation}
where $\Pi$ is the matrix containing the Goldstone degrees of freedom. 
The latter decompose under the SM gauge group as
$\mathbf{1_0 \oplus 3_0 \oplus 2_{1/2} \oplus 3_1}$ and 
are given by
\begin{equation}
\Pi = \left(\begin{array}{ccc}
0_{2\times 2} & \frac{H}{\sqrt{2}} & \Phi  \\
\frac{H^\dagger}{\sqrt{2}} & 0 & \frac{H^T}{\sqrt{2}}  \\
\Phi^\dagger & \frac{H^*}{\sqrt{2}} & 0_{2\times 2}  \\
\end{array}\right) \,. \label{pionmat}
\end{equation}
Here, $H= (-i\pi^+, \frac{h+i\pi^0}{\sqrt{2}})^T$ is the $SU(2)$ Higgs
doublet $\mathbf{2_{1/2}}$ and $\Phi$ is the complex triplet
$\mathbf{3_1}$ which forms a symmetric tensor with components
$\phi^{\pm\pm},\phi^\pm,\phi^0,\phi^P$.  After EWSB, $\pi^+$ and
$\pi^0$ will be eaten by the SM gauge bosons $W$ and $Z$.  The
invariance of the Lagrangian under $T$-parity demands the scalar to
transform as
\begin{equation}
T: \Pi \rightarrow -\Omega \, \Pi \, \Omega \qquad \Longrightarrow \qquad 
\Sigma \rightarrow \Sigma_0 \, \Omega \, \Sigma^\dagger \, \Omega \, 
       \Sigma_0  \ ,
\end{equation}
with $\Omega = \rm diag(1,1,-1,1,1)$.  The transformation rules
guarantee that the complex triplet field is odd under $T$ parity,
while the (usual) Higgs doublet is even.  This has the consequence
that the SM gauge bosons do not mix with the $T$-odd heavy gauge
bosons, thereby prohibiting any further corrections to the low energy
EW observables at tree level and thus relaxing the EW constraints on
the model~\cite{Hubisz:2005tx}.

After the electroweak symmetry breaking, the masses of the $T$-odd
partners of photon($A_h$), $Z$ boson($Z_h$) and $W$ boson($W_h$) are
given by,
\begin{equation}
\label{gauge-mass}        
\begin{array}{rcl}
M_{A_{h}}  &\simeq& \displaystyle \frac{g^\prime f}{\sqrt{5}} \,
             \left(1 - \frac{5v^2}{8f^2}\right)\,,  \\[2ex]
M _{Z_{h}} \simeq  M_{W_{h}}  & \simeq &  \displaystyle
g f \left(1 - \frac{v^2}{8f^2} \right)\ ,
\end{array}
\end{equation}
with $v \simeq$ 246 GeV being the electroweak breaking scale.  The
heavy photon $A_h$ is the lightest $T$-odd particle (LTP) 
and can
serve as the DM candidate with the correct relic
density\cite{Birkedal:2006fz, Asano:2006nr, Chen:2014wua}.

Implementation of $T$-parity in the fermion sector requires a doubling
of content and each fermion doublet of the SM must be replaced by a
pair of $SU(2)$ doublets $(\Psi_1,\Psi_2)$. Under $T$-parity, the
doublets exchange between themselves $(\Psi_1 \leftrightarrow \Psi_2)$
and  the $T$ even combination remains almost massless and is identified with
the SM doublet. On the other hand, the $T$ odd combination
  acquires a large mass\footnote{A recent study of the heavy top partner production 
  	at the LHC including the global analysis of this model has been done in~\cite{Han:2014qia,Liu:2015kmo}.}, courtesy a Yukawa coupling involving the
  large vev and an extra $SU(2)$ singlet fermion (necessary, anyway,
  for anomaly cancellation). 
 For simplicity, we can assume an universal
and flavor diagonal Yukawa coupling $\kappa$ for both up and down type
fermions. The mass terms will then, respectively, be
\begin{equation}
\begin{array}{rcl}
M_{d_h} &\simeq& \sqrt{2} \kappa f\,, \\
M_{u_h} &\simeq& \displaystyle
\sqrt{2} \kappa f \, \left(1 - \frac{v^2}{8f^2} \right)\,.
\end{array}
\end{equation}

If $f \sim {\cal O}({\rm TeV})$, the masses for the exotic up and down
type fermions become comparable. Since our study concentrates on the first two
generations of $T$-odd heavy fermions,  
we desist from a discussion of the top sector and 
point the reader
to Refs.~\cite{Low:2004xc,Hubisz:2004ft,Hubisz:2005tx}.  
Thus, in a nutshell, the phenomenology relevant to this paper is 
characterized by only two parameters, the scale $f$ and the universal
Yukawa coupling $\kappa$.

\section{Numerical Analysis }\label{analysis} 
We now present a detailed discussion of our analysis, which pertains
to the case of large $\kappa$, or, in other words the situation where
the $T$-odd fermions are significantly heavier than the $T$-odd gauge
bosons.  We limit ourselves to a study of the dominant processes,
viz. the production of a pair of such fermions (antifermions) on the
one hand, and the associated production of a heavy gauge boson
alongwith one such fermion. In other words, the processes of interest
are:
\begin{subequations}
	\begin{eqnarray}
	(a)~ p~ p &\rightarrow& Q_{h_i}~{\bar Q_{h_j}}, \quad 
                             Q_{h_i} Q_{h_j}, \quad  {\bar Q_{h_i}} {\bar Q_{h_j}} \label{s1} \\
	(b)~ p~ p  &\rightarrow& Q_{h_i}/{\bar Q_{h_i}}~ W_h^{\pm} \label{s2}\\
	(c)~ p~ p &\rightarrow& Q_{h_i}/{\bar Q_{h_i}}~ Z_h \label{s3}
	\end{eqnarray}	\label{processes}
\end{subequations}
where $Q_{h_i}, Q_{h_j}, (i,j=1,2)$ denote the first two generations
of heavy $T$-odd quarks $(u_h,d_h,c_h,s_h)$, whereas
$W_h^{\pm}$ and $Z_h$ are the $T$-odd heavy partners of the SM $W$-boson
and $Z$-boson respectively. We
  focus mainly on the current and future runs of the LHC, keeping in
  mind the constraints on the parameter space ensuing from the
  negative results of Run I (center of mass energy $\sqrt{s} = 8$
  TeV)~\cite{Reuter:2013iya}.  
\begin{table}[htbp!]
\vskip 5pt
	\centering
	\begin{tabular}{|p{2cm}|c|c|c|c|c|c|}
		\hline
		Benchmark Points & f (GeV) & $\kappa$ & $M_{u_h}~(\rm GeV)$ & $M_{d_h}~(\rm GeV)$ & $M_{W_h}=M_{Z_h}~(\rm GeV)$ & $M_{A_h}~(\rm GeV)$ \\ \hline
		BP1 & 1100& 0.8 & 1236.8& 1244.5& 712.3& 170.4 \\ \hline
		BP2 & 1200& 0.75 & 1266 & 1273 & 778.2 & 186.8 \\ \hline
	\end{tabular}
	\caption{\it Benchmark points and corresponding masses of the $T$-odd particles}
	\label{bp}
\vskip 5pt
\end{table}
Rather than presenting a scan over the parameter space, 
we choose two representative benchmark points (consistent with the 
present constraints) that illustrate not only the sensitivity of the 
experiments to the two-dimensional parameter space ($f$,$\kappa$), 
but also the bearing that the spectrum has on the kinematics and, hence, 
the efficiencies. In Table~\ref{bp}, we
list the values of the scale $f$ and the Yukawa coupling $\kappa$ for
the chosen benchmark points (BP), as also the relevant part of the $T$-odd spectrum.
The corresponding branching ratios of the up-type heavy  
quarks $(u_{hi})$ and the heavy gauge bosons are 
\begin{subequations}
	\begin{eqnarray}
	{\rm BR} (u_h \to W^+_h ~d) & \approx& 60\% \nonumber \\
	{\rm BR} (u_h \to Z_h ~u)  & \approx& 30\% \nonumber \\
	{\rm BR} (u_h \to A_h ~u)  & \approx& 10\% \\
	{\rm BR} (W^+_h \to W^+ ~A_h) &\approx& 100\% \\
	{\rm BR} (Z_h \to H ~A_h)  &\approx& 100\% \label{dk}
	\end{eqnarray}
\end{subequations}
where $H$ is the light (standard model-like) Higgs. 
The branching ratios for the down-type heavy quarks
$(d_{hi})$ are very similar to those for $u_{hi}$.  
Furthermore, with the available phase space being quite large in each
case, the kinematic suppression is negligible. Consequently, there is
relatively little difference between the branching ratios (less than
0.5\%) for the two benchmark points. And, while, for more extreme points,
the difference could be slightly larger, the situation does not change
qualitatively.

The three sub-processes of Eq.~(\ref{processes}) can, thus, give rise
to the following three possible final states\footnote{
	Of several possibilities, we concentrate only on final states 
	with leptons. This not only ensures a good sensitivity, but is also, 
	experimentally, very robust and least likely to suffer on account of
	the level of sophistication of our analysis.  However, 
	non-leptonic final states may also provide interesting signal
	topologies. For example, hadronic decays of $W/H$, with their larger
	branching ratios as well as di-higgs final state where both the Higgs
	decay to $b \bar b$ channel can be studied exploiting jet substructures. 
	Such a all-encompassing analysis is, though, beyond the scope of this
	paper.}$^,$\footnote{{Note that 
final states with additional charged leptons are also possible, but 
the corresponding branching fractions are smaller. Thus, the signal
size is likely to prove a bottleneck in spite of a possibly better 
discriminatory power.}}:
\begin{eqnarray}	
 &(i)&	1\ell^\pm + nj + \mET \qquad   n \geq 3 \,. \nonumber \\
 &(ii)&	1\ell^\pm + 2b + j + \mET \,. \nonumber \\
 &(iii)&	2\ell^\pm + nj + \mET \qquad n \geq 2  \,. 
\label{final_states}
\end{eqnarray}
where, $\ell = e, \mu $; $b$ corresponds to a $b$-tagged jet and $j$ denotes non
$b$-tagged jets.  The leading order (LO) production cross-sections
for each of the sub-processes listed in Eq.~(\ref{processes}) are 
calculated using {\tt MadGraph5} ~\cite{Alwall:2014hca} and are listed 
in Table~\ref{cs}, wherein we have used the 
{\tt Cteq6L} parton distributions.  Since the $K$-factors are larger 
than unity, the use of the LO cross sections for the signal 
events is a conservative choice. The larger production cross-sections
for BP1 (as compared to BP2) is but a consequence of the lighter 
masses for the exotic particles.
\begin{table}[htbp!]
\vskip 5pt
	\centering
	\begin{tabular}{|p{2cm}|c|c|c||c|c|c|}
		\hline
		Benchmark & \multicolumn{3}{|c||}
                  { Production cross-section} &  \multicolumn{3}{|c|}{Production cross-section}  \\ 
 Points  &  \multicolumn{3}{|c||}{(in fb) at $\sqrt{s} = 13$ TeV} &
 \multicolumn{3}{|c|}{(in fb) at $\sqrt{s} = 14$ TeV} \\ \cline{2-7}
		& Process $(a)$ & Process $(b)$ & Process $(c)$ & Process $(a)$ & Process $(b)$ & Process $(c)$  \\ \hline
		BP1  & 129.1 & 51 & 25 & 172.3 & 68 & 34 \\ \hline
		BP2  & 97 & 37 & 19 & 131.9 & 50.5 & 25.3 \\ \hline
	\end{tabular}
	\caption{\it Production cross sections for the various processes at 13 and 14 TeV LHC.}
	\label{cs}
\vskip 5pt
\end{table}
For our analysis, we use {\tt Madgraph5} to generate the events at
parton level at LO for both the signal as well as the SM background
contributing to the respective final states under consideration.  The
model files for LHT, used in {\tt Madgraph5} are generated using
FeynRules~\cite{Alloul:2013bka}\hldg{\footnote{We thank the authors of
    Ref.~\cite{Reuter:2013iya} for sharing the UFO model files.}}.
The unweighted parton level events are then passed for showering
through {\tt Pythia}(v6.4)~\cite{Sjostrand:2006za} to simulate
showering and hadronisation effects, including fragmentation. For
Detector simulation, we then pass these events through {\tt
  Delphes}(v3)~\cite{deFavereau:2013fsa} where jets are constructed
using the anti-$k_T$ jet clustering algorithm with proper MLM matching
scheme chosen for background processes.  Finally, we perform the cut
analyses\footnote{It is reassuring to note that even a parton model
  analysis reproduces much of the results at the 10\% level, which, in retrospect, is 
  not surprising given the relatively clean nature of the signal and the presence 
of the lepton and missing energy.} using {\tt
  MadAnalysis5}~\cite{Conte:2012fm}.

Several SM sub-processes constitute backgrounds to the
aforementioned final states. In particular, one needs to consider:
\begin{itemize}
\item $t\bar t (+ \rm jets)$: Comprising the semi-inclusive cross-section for $t
  \bar t$ production with up to two additional hard jets, this
  constitutes the dominant background for all the three final
  states. For example, the orders of magnitude larger cross section
  for top-production means that a disconcertingly large number of such
  events would satisfy the requirement of a pair of $b$-jets
  reconstructing to the SM Higgs peak.  

\item $W^\pm + \rm jets$: With a significantly hard $ \mET$ 
distribution, this process serves as the dominant
  background for the signal configuration with a single charged lepton in the final
  state (and no $b$-jets).  We consider here, the 
  semi-inclusive cross section
    for $W^\pm$ with up to three hard jets.

\item $Z + \rm jets$: While this could have been
  the major background for the signal configuration
  with two charged leptons in the final state,
  a large $\mET$ requirement can effectively 
  suppress it. Akin to the case for the $W^\pm+jets$ background, this too
    includes the semi-inclusive cross section for the production of $Z$
    with up to three hard jets.

\item Diboson $+ \rm jets$:  With large production
cross-sections, $WW~(WZ,ZZ)$ with two hard jets production in SM 
are significant sources of background. For example, owing to mismeasurements,
a $b \bar b$ pair from a $Z$-decay could fake a Higgs. In addition, 
mistagging constitutes another source for such backgrounds. 

\item Single top production: This will contribute mainly to final state $(i)$.

\item $t\bar t (+ W/Z/H)$: Similar to $t\bar t (+ \rm jets)$, 
these processes may also contribute to the total SM background,
but with much lower production cross-sections. 

\end{itemize} 

Since the final states under discussion can also result from 
hard subprocesses accompanied by 
 either or both of initial and final state radiation, or  soft decays, we 
must impose some basic cuts before we attempt to simulate the events. 
To this end, we demand that
\begin{subequations}
\begin{eqnarray}
\Delta R_{\ell i} > 0.2\,, && \Delta R_{ij} > 0.7\,, \\
\Delta R_{bi} > 0.7\,, &&  (i = \ell,j,b) \\ 
p_T^j > 30 ~{\rm GeV}\,, && |\eta_j| < 5\\
p_T^\ell > 5 ~{\rm GeV}\,, && |\eta_\ell| < 2.5
\,. 
\end{eqnarray}\label{basic_cuts}
\end{subequations}
 Following the ATLAS collaboration~\cite{ATLAS-CONF-2014-046}, we consider a $p_T$-dependent
$b$-tagging efficiency as below:
\begin{eqnarray}
\epsilon_b = \begin{cases}
0 & p_T^b \leq 30 ~\rm GeV \\
0.6 & 30 ~{\rm GeV} < p_T^b < 50 ~{\rm GeV} \\
0.75 & 50 ~{\rm GeV} < p_T^b < 400 ~{\rm GeV} \\
0.5 & p_T^b > 400 ~\rm GeV 
\end{cases} 
\label{b-tag}
\end{eqnarray}
Along with this, we also incorporate a mistagging probability of 10\% (1\%) for charm-jets (light-quark and gluon jets). Also, the absolute rapidity of $b$-jets are demanded to be less than 2.5 $(|\eta_b| < 2.5)$.

\begin{figure}[ht!]
	\centering
	\begin{tabular}{cc}
		\includegraphics[height=4cm,width=8cm]{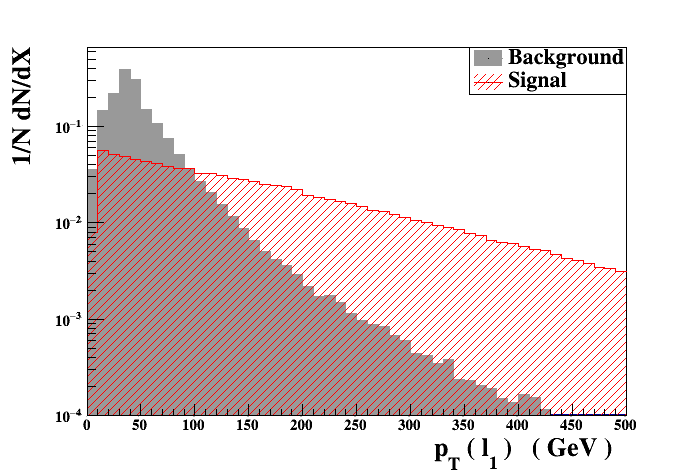} & \includegraphics[height=4cm,width=8cm]{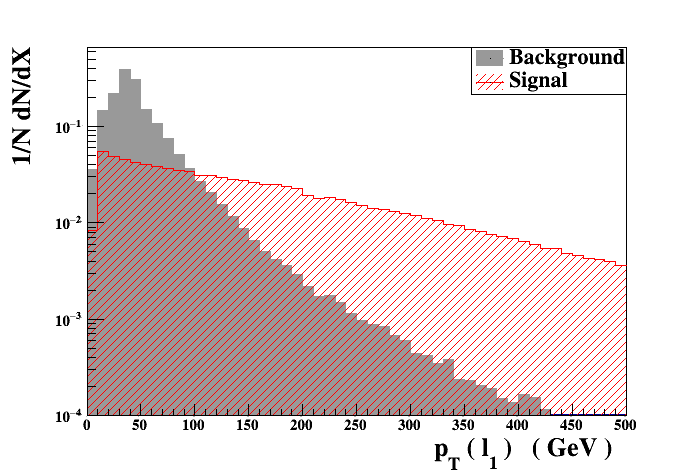} \\
		\includegraphics[height=4cm,width=8cm]{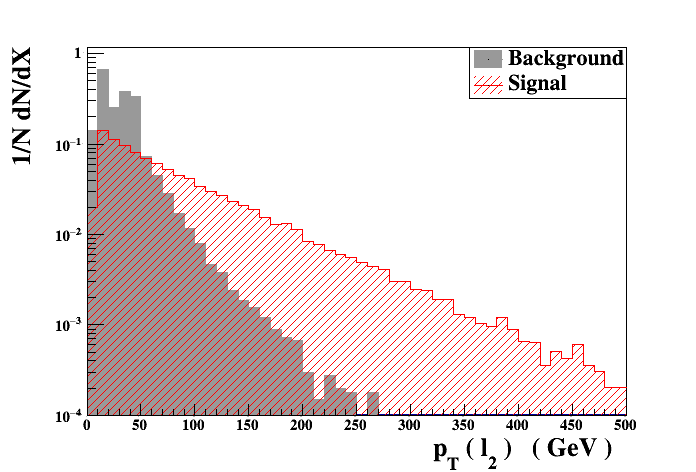} & \includegraphics[height=4cm,width=8cm]{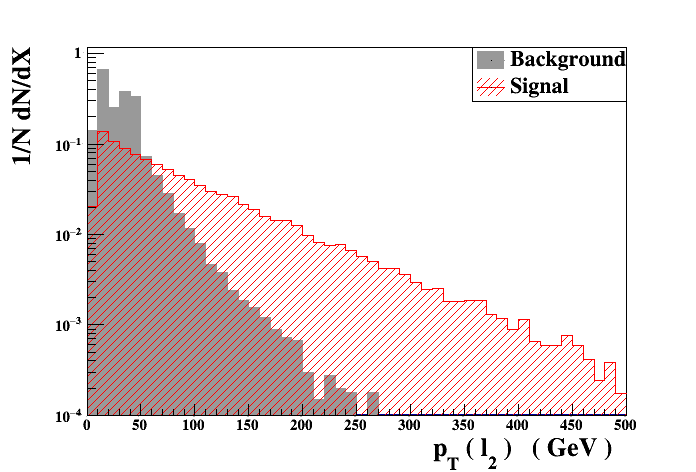} 
	\end{tabular}
	\caption{\it Normalized $p_T$ distributions 
          for the leading (upper panels) and subleading (lower panels) 
            leptons. The left and right panels refer to 
            BP1 and BP2 respectively.
}
	\label{ptl}
\end{figure}

\begin{figure}[ht!]
  \centering
  \begin{tabular}{cc}
    \includegraphics[height=4cm,width=8cm]{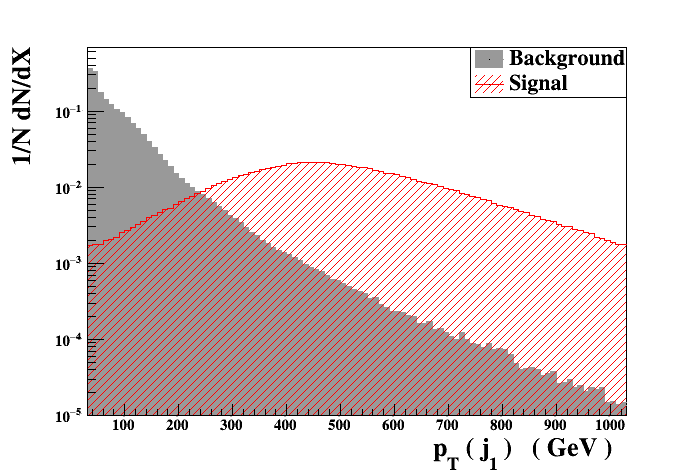} 
  & \includegraphics[height=4cm,width=8cm]{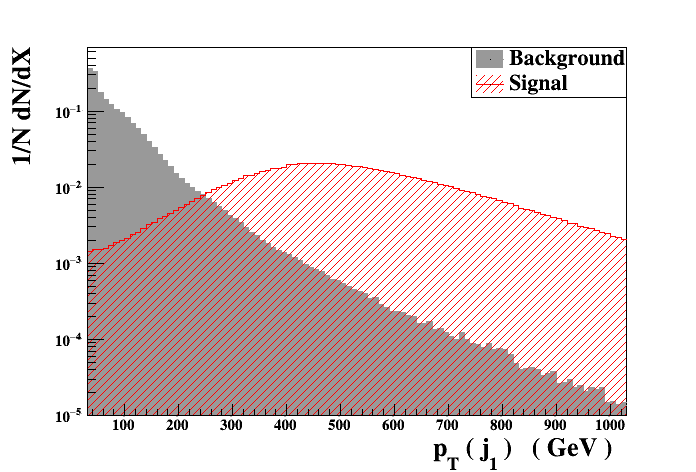} \\
    \includegraphics[height=4cm,width=8cm]{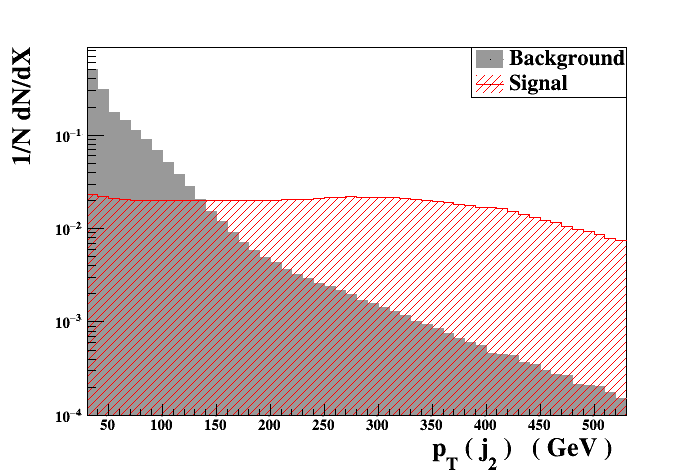} 
  & \includegraphics[height=4cm,width=8cm]{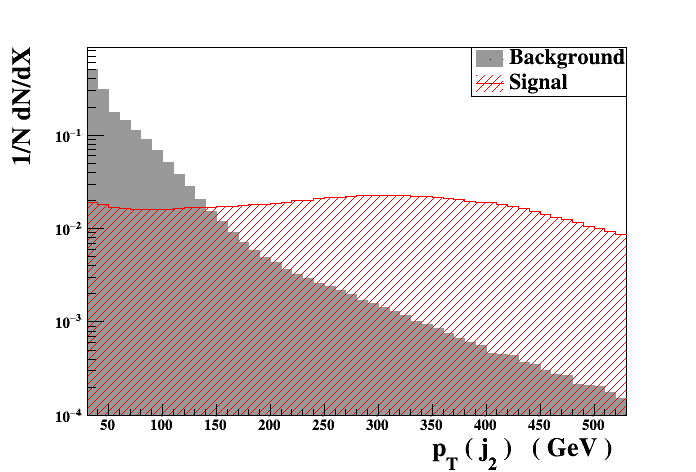} \\	
    \includegraphics[height=4cm,width=8cm]{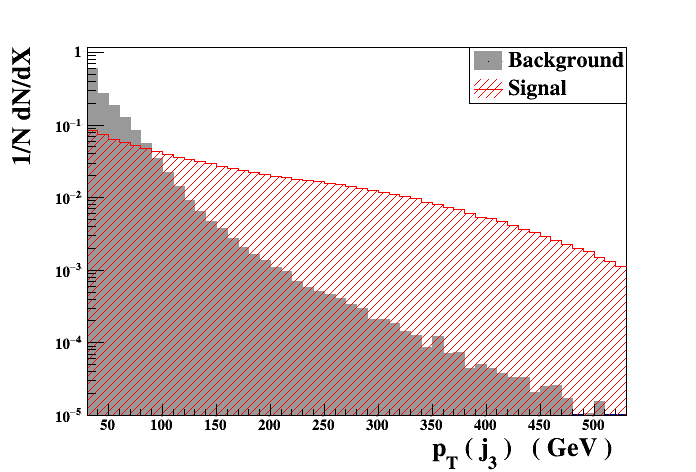} 
  & \includegraphics[height=4cm,width=8cm]{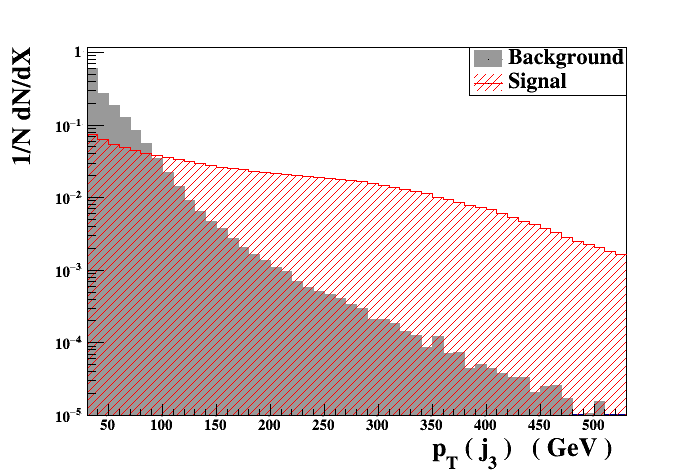} 
  \end{tabular}
  \caption{\it Normalized $p_T$ distributions              
    for the three leading (top
    to bottom) non $b$-tagged jets for BP1 (left  
    sets) and BP2 (right sets).}
  \label{ptj}
\end{figure}

\begin{figure}[ht!]
  \centering
  \begin{tabular}{cc}
    \includegraphics[height=4cm,width=8cm]{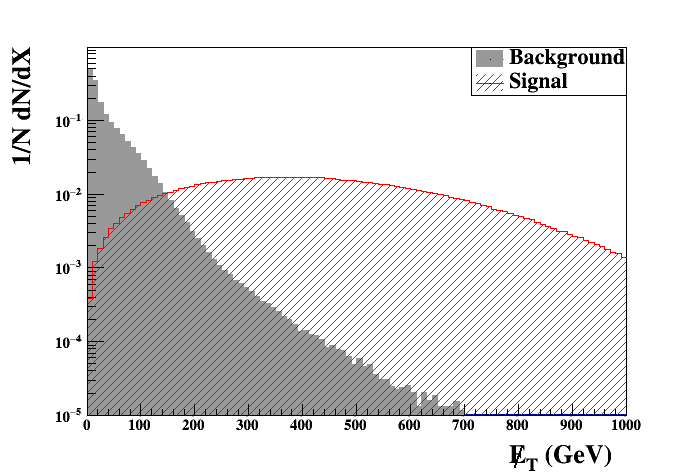} 
  & \includegraphics[height=4cm,width=8cm]{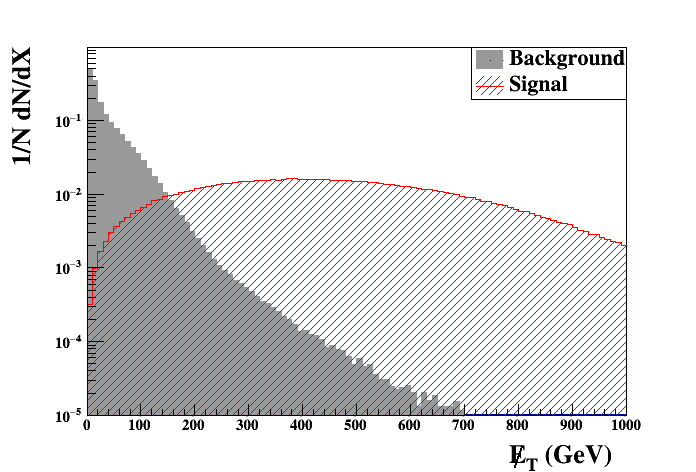} \\
    \includegraphics[height=4cm,width=8cm]{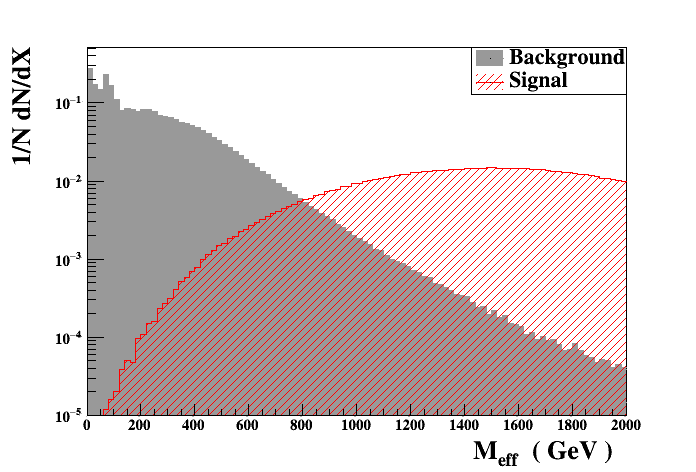} 
  & \includegraphics[height=4cm,width=8cm]{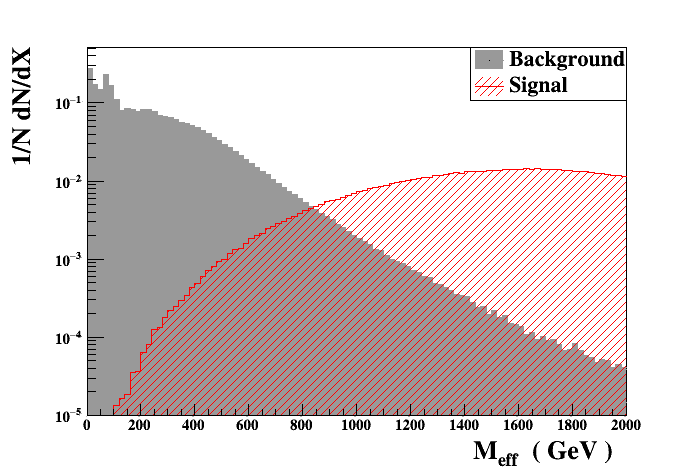} \\
    \includegraphics[height=4cm,width=8cm]{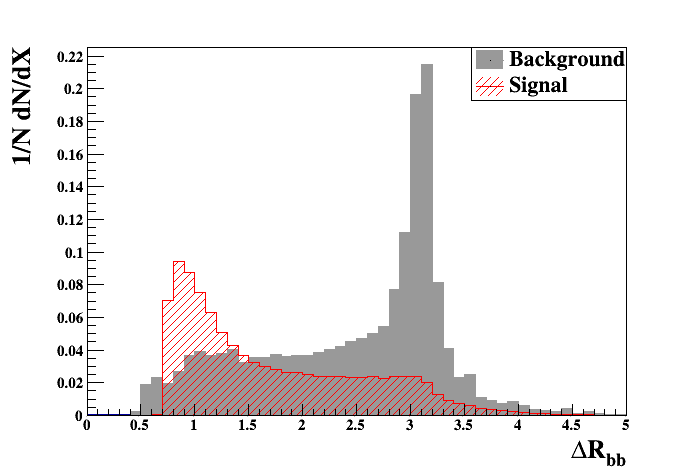} 
  & \includegraphics[height=4cm,width=8cm]{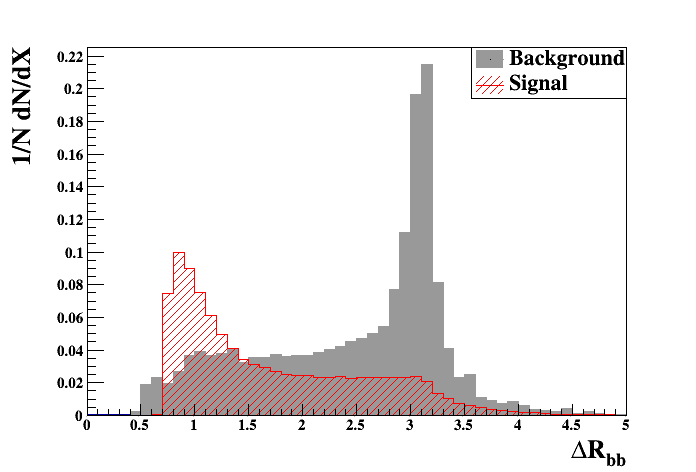} \\
    \includegraphics[height=4cm,width=8cm]{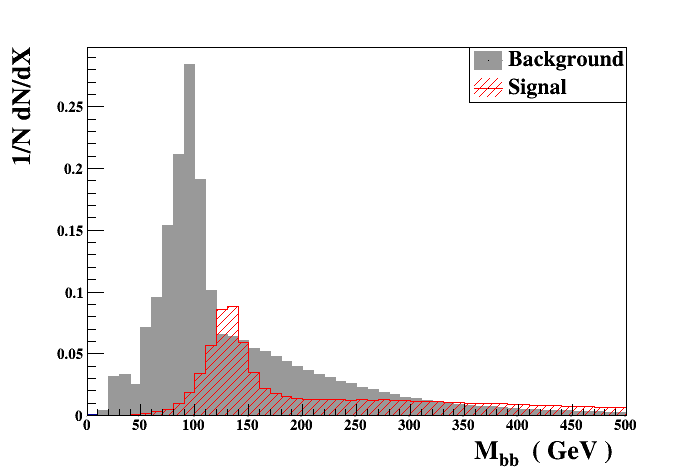} 
  & \includegraphics[height=4cm,width=8cm]{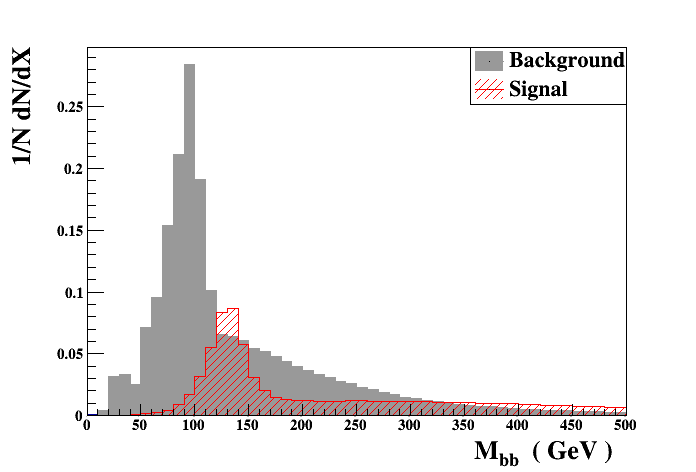} 
  \end{tabular}
  \caption{\it Normalized distributions 
    for the missing transverse energy (top panels), the effective 
    mass (second row), 
    the $b\bar b$ invariant mass (third row) 
    and ${\Delta R}_{\rm bb}$ (bottom panels). In each 
case, the left and right panels correspond to BP1 and BP2 respectively.
  }
	\label{met}
\end{figure}

We show, in Figs.~\ref{ptl},\ref{ptj} and \ref{met},
the histograms for the signal and background events after imposing only the basic cuts of
Eq.~(\ref{basic_cuts}).

To understand the transverse momentum distribution of the leading
lepton (the upper panels of Fig.~\ref{ptl}), recall the
decay chain for the signal processes.  In all of the three processes,
the heavy $W_h^\pm$ is produced, either directly or from the decay of
heavy $T$-odd quarks.  As already mentioned, for the parameter space
of interest, the $W_h^\pm$ decay to $W^\pm+A_h$ with almost 100\%
branching ratio and, hence, the subsequent decay of the $W^\pm$
generates leptons in the final state. With the mass difference between
the $W_h^\pm$ and SM $W^\pm$ bosons being so large, the latter would,
typically, have a large $p_T$, even if the former had a small
$p_T$. This translates to a large $p_T$ for the charged lepton
emanating from the $W^\pm$ decay. Thus, for most events resulting from
the process of (\ref{s2}), the leading lepton tends to have a large
$p_T$. For the other two production channels, at least a large
fraction of the events would have the $Q_h$ decaying into $W_h^\pm$,
thereby bestowing the latter with a large $p_T$ to start with. It
should be realized though, that in each case, the possibility exists
that, in a decay, the $p_T$ of a daughter, as defined in mother's rest
frame, is aligned against the mother's $p_T$. While this degradation
of the $p_T$ is not very important for the leading lepton, it
certainly is so for the next-to-leading one, as is attested to by the
lower panels of Fig.~\ref{ptl}. It is instructive to examine the
  corresponding distributions for the background events. Since the
  $W$'s (or $Z$'s) now have typically lower $p_T$, the Jacobian peak
  at $m_W/2$ ($m_Z/2$) is quite visible, and particularly so for the
  next-to-leading lepton. For the leading one, the peak,
  understandably, gets smeared on account of the inherent $p_T$ of the
  decaying boson. This effect, of course, is more pronounced for the
  signal events. The second, and more pronounced, peak in the lower
  panels of Fig.~\ref{ptl} result from non-resonant processes and/or
  configurations wherein the lepton travels against the direction of
  its parent. This motivates our cuts on the lepton $p_T$s.

In an analogous fashion, the decay of the heavy
$T$-odd quark (almost) always yields a high-$p_T$ 
jet owing to the large difference between its mass and those 
for $T$-odd bosons. Consequently, a requirement of $p_T(j_1) \gtrsim 250$~GeV 
would eliminate only a very small fraction of the signal events (for each of the 
three channels, while, potentially, removing a significant fraction of the background 
events (see upper panels of Fig.~\ref{ptj}). 
For process (\ref{s1}), the decay of the second $Q_h$ would lead 
to a jet almost as hard. On the other hand, for channels (\ref{s2} \& \ref{s3}),
the second jet results only from the decays of the $W^\pm$ or the 
$H$. Intrinsically much softer, these still gain from the $p_T$ 
of the mother. Consequently, the second leading jet, very often, may have 
a $p_T$ larger than 200 GeV (see the middle panels of Fig.~\ref{ptj}).
For the processes under discussion, a third jet can only result from the (cascade) decay 
of a SM boson, and, hence, is typically softer (the lower panels of Fig.~\ref{ptj}) and hence, 
for three-jet final states, the requirement on the third jet 
$p_T$ should not be much stricter than about 45~GeV.

Next, we turn to some derived kinematic variables. The first 
quantity of interest is the missing transverse energy ($\mET$).
For the background events, this arises from the 
neutrinos (courtesy $W^\pm$ or $Z$ decays) or mismeasurement of 
the jet and lepton momenta. 
For the signal events, this receives an additional contribution 
from the heavy photon $A_h$, which is stable because of $T$-parity. 
Not only does the $A_h$ have a large mass, but also a large $p_T$ 
owing to the large difference in the masses of the mother and the 
daughters (on each occasion wherein it is produced). Consequently,
the $\mET$ spectrum is much harder for the signal (the upper panels 
of Fig.~\ref{met}) and a requirement of $\mET > 250$~GeV would 
significantly improve the signal to background ratio.

Another variable of interest is the effective mass variable defined as
\begin{equation}
M_{\rm eff} = \sum_{n=1}^{4}|p_T(j)|_n + \mET \,
\end{equation}
where the sum goes over up to four jets.  Similar to the case for the
$\mET$ distribution, the high masses of the exotic $T$-odd particles
leads to a large $M_{\rm eff}$ for the signal events as one can see
from the second panel of Fig.~\ref{met}. Hence, a significant ($\rm
M_{eff}$) cut also helps in reducing the background events.

Finally, for events with two tagged $b$-jets, we may consider
  the invariant mass of the pair (the third row of Fig.~\ref{met}).
In the signal events, the $Z_h$ decays to a Higgs boson and
$A_h$, with the former decaying
predominantly into a $b {\bar b}$ pair. Due to the
large mass difference between $Z_h$ and $A_h$, the Higgs boson will be
produced with a high $p_T$ which will be imparted to its decay products. 
As a result, the $b {\bar b}$ pair will be produced with a relatively small
opening angle. On the other hand, the $b$'s in the SM background arise, 
primarily, from three classes of processes: ($a$) the decay of 
different top-quarks where the separation between them show a much 
broader structure; ($b$) the decay of a $Z$ boson, wherein the invariant mass 
would peak at $m_Z$, and owing to the relatively low momentum of the $Z$, the 
$b$'s be well-separated (in fact, close to being back-to-back); and 
($c$) from non-resonant processes where the $b$'s would be softer and, again, 
$\Delta R_{b\bar b}$ would have a wider distribution. These features are well
reflected by the third and fourth rows of Fig.~\ref{met}.  
It is, thus, expected that a judicious upper cut on
$\Delta R_{b \bar b}$ would definitely improve the signal
significance. Similarly, a good energy-momentum resolution for
the $b$-jets would serve to remove much of the $Z$-background.

\subsection{Cut Analysis:}
All the processes in Eq.~(\ref{processes}) may contribute to 
a given final state---of Eq.~(\ref{final_states})---and, henceforth,
we include all under `Signal', while `Background' receives contributions
from all SM processes leading to the particular final state.

In addition to the basic cuts of Eq.~(\ref{basic_cuts}), further
selection cuts may be imposed in order to {improve the signal to background 
ratio. Understandably, these selections cuts would depend on the 
final state under consideration, both in respect of the differences 
in event topology for the signal and the background, as well as 
on the actual size of the signal. In particular, we are guided by the 
requirement that not too large an integrated luminosity is required 
to reach a $5\sigma$ significance ($S = 5$), with $S$ being}  
\begin{eqnarray}
S &=& \frac{N_S}{\sqrt{N_S+N_B}} \, \ ,
\label{significance}
\end{eqnarray}
{where} $N_S$ and $N_B$ represent number of signal and background
events respectively. We now take up each final state given in Eq.
(\ref{final_states}) and describe the kinematic cut flow followed in
selecting events for the signal while suppressing the background.

\subsubsection{$1\ell^\pm + nj + \mET$ ; $n \geq 3 $}
\label{subsec:i}
This final state for the signal receives 
contribution from both the
strongly produced $T$-odd quark pair as well as the associated
production modes, thus giving us the maximum signal event rate amongst
the final states under consideration. Here the single charged lepton
almost always comes from the decay of a $W$ boson resulting from the
cascades.  The selection cuts, in the order that they are imposed,
are:
\begin{enumerate}
\item $p_T(j_3) > 45$ GeV and $|\eta(j)|< 2.5$ (C1--1): In other
words, we demand that our final state has at least three jets
within the given pseudo-rapidity range, each with a minimum
$p_T$ of 45 GeV. This choice is motivated by the lowest
panels of Fig.~\ref{ptj}.
	
\item $p_T(j_1) > 250$ GeV (C1--2): Given that the hardest jet is, typically, 
 much harder for the signal events than it is for the background (see 
 top panels of Fig.~\ref{ptj}), we ask that $p_T(j_1) > 250$~GeV. This, 
 understandably, helps increase the signal to noise ratio to a remarkable 
 extent.
	
\item $p_T(j_2) > 200$ GeV (C1--3): Similarly, we demand that the
  $p_T$ of the second hardest jet be more than 200 GeV. In
  particular, this reduces (moderately) the $t\bar t$ and $W + n$-jets
  background (with only a smaller effect on the $Z$ background) while
  having only a marginal effect on the signal strength. 
  
\item $\rm M_{eff} > 1.2$ TeV (C1--4): Motivated by the second row of Fig.~\ref{met}, 
we demand that $M_{\rm eff} > 1.2$ TeV.  Such a high value effectively suppresses the 
background without reducing much of the signal.
	
\item $\mET > 250$ GeV (C1--5): The top row of Fig.~\ref{met} bears 
  our (previously stated) expectations that the extent of transverse 
  momentum imbalance ($\mET$) would be far larger for the signal events than that for the background. Consequently, this requirement improves considerably the 
  signal to background ratio.
	
\item $p_T(\ell_1) > 20$ GeV (C1--6): Finally, to distinguish this final
state from that considered in Sec.~\ref{subsec:iii}, we require that there be
  {\em only one} isolated charged lepton $e,\mu$ with a $p_T$ more than 
  20 GeV. 	
\end{enumerate}

\def\nl{\cline{1-8}&&&&&&&&\\[-2.5ex]}
\begin{table}[ht!]
	\centering
	\footnotesize
	\begin{tabular}{|p{1.7cm}|p{2.1cm}|c|c|c|c|c|c|c|}
		\hline
		\multicolumn{2}{|c|}{}& \multicolumn{6}{|c|}{Effective Cross-section(fb) after the cut} & ${\cal L}_{5\sigma}$\\ \hline
		SM-background & Production Cross-sec. (fb) 
		& C1--1 & C1--2 & C1--3 & C1--4 & C1--5 & C1--6 &  
		  \\  \nl 
		t $+$ jets & $2.22 \times 10^5$ & \t{6.12}{4} & \t{3.34}{3} & \t{1.84}{3} & 202 & 13.8 & 2.8  & \\  \nl 
		t$\bar{\rm t} +$ jets & $7.07 \times 10^5$ & \t{4.37}{5} & \t{4.54}{4} & \t{2.51}{4} & \t{5.38}{3} & 453 & 123 & \\  \nl
		W $+$ jets & $1.54 \times 10^8$ & \t{1.37}{6} & \t{9.06}{4} & \t{5.33}{4} & \t{8.99}{3} & 679 & 39.9 &  \\ \nl
		Z $+$ jets & $4.54 \times 10^7$ & \t{8.49}{5} & \t{4.12}{4} & \t{2.65}{4} & \t{3.75}{3} & 162 & 0 &  \\  \nl
		WW $+$ jets & $8.22 \times 10^4$ & \t{3.50}{4} & \t{1.42}{3} & 800 & 51.2 & 5.1 & 1.5 &  \\  \nl
		ZZ $+$ jets & $1.10 \times 10^4$ & \t{1.91}{3} & 162 & 103 & 21.9 & 2.8 & 0.2  & \\  \nl
		WZ $+$ jets & $3.81 \times 10^4$ & \t{8.30}{3} & \t{1.39}{3} & 888 & 214 & 29.4 & 5.6  & \\ \nl
		t$\bar{\rm t}+$W & 351 & 238 & 45.8 & 24.9 & 5.9 & 0.8 & 0.2 &  \\  \nl
		t$\bar{\rm t}+$Z & 585 & 435 & 101 & 55.3 & 15.6 & 1.8 & 0.3 &  \\  \nl
		t$\bar{\rm t}+$H & 400 & 316 & 69.9 & 39.8 & 11.6 & 1.1 & 0.3 &  \\  \cline{1-8}
		Total background & \multicolumn{6}{|c|}{ } & 173.8 &   \\ \hline \hline
		BP1 & 205 & 168 & 164 & 148 & 138 & 114 & 25.3 & 7.8  $\rm fb^{-1}$ \\ \hline
		BP2 &  153 & 129 & 126 & 115 & 109 & 91.8 & 15.2 & 20.4  $\rm fb^{-1}$ \\  \hline
	\end{tabular}
	\caption{\it Cut-flow for final state $(i)$ at $\sqrt{s}= 13$ TeV. 
${\cal L}_{5\sigma}$ is the integrated luminosity required 
       to reach a  $5\sigma$ significance.}
	\label{cut_anal1_13}
\end{table}

\begin{table}[ht!]
	\centering
	\footnotesize
	\begin{tabular}{|p{1.7cm}|p{2.1cm}|c|c|c|c|c|c|c|}
		\hline
		\multicolumn{2}{|c|}{}& \multicolumn{6}{|c|}{Effective Cross-section(fb) after the cut} & ${\cal L}_{5\sigma}$\\ \hline
		SM-background & Production Cross-sec. (fb) 
		& C1--1 & C1--2 & C1--3 & C1--4 & C1--5 & C1--6 &  
 	\\  \nl
		t $+$ jets & $2.49 \times 10^5$ & \t{6.86}{4} & \t{3.74}{3} & \t{2.06}{3} & 226 & 15.5 & 3.1  & \\  \nl
		t$\bar{\rm t} +$ jets & $7.96 \times 10^5$ & \t{4.92}{5} & \t{5.11}{4} & \t{2.82}{4} & \t{6.06}{3} & 509 & 138  & \\  \nl
		W $+$ jets & $1.66 \times 10^8$ & \t{1.47}{6} & \t{9.69}{4} & \t{5.71}{4} & \t{9.63}{3} & 728 & 42.8  & \\   \nl
		Z $+$ jets & $4.86 \times 10^7$ & \t{9.10}{5} & \t{4.42}{4} & \t{2.84}{4} & \t{4.03}{3} & 173 & 0 &  \\  \nl
		WW $+$ jets & $9.04 \times 10^4$ & \t{3.85}{4} & \t{1.56}{3} & 879 & 56 & 5.6 & 1.7  & \\ \nl
		ZZ $+$ jets & $1.18 \times 10^4$ & \t{2.04}{3} & 173 & 109 & 23.4 & 3.1 & 0.2 &  \\  \nl
		WZ $+$ jets & $4.21 \times 10^4$ & \t{9.18}{3} & \t{1.54}{3} & 980 & 236 & 32 & 6.2 &  \\ \nl 
		t$\bar{\rm t}+$W & 398 & 270 & 52.0 & 28.3 & 6.7 & 0.9 & 0.3 &  \\  \nl
		t$\bar{\rm t}+$Z & 706 & 526 & 122 & 66.8 & 18.8 & 2.2 & 0.4 &  \\  \nl
		t$\bar{\rm t}+$H & 479 & 379 & 83.7 & 47.7 & 13.9 & 1.3 & 0.4 &  \\  \cline{1-8}
		Total background & \multicolumn{6}{|c|}{ } & 193.1 &  \\ \hline \hline 
		BP1 & 275 & 225 & 220 & 198 & 186 & 154 & 34.0 & 4.9 $\rm fb^{-1}$ \\ \hline
		BP2 &  208 & 174 & 170 & 156 & 148 & 124 & 20.7 & 12.5  $\rm fb^{-1}$ \\  \hline
	\end{tabular}
	\caption{\it Cut-flow for final state $(i)$ at $\sqrt{s}= 14$ TeV.}
	\label{cf1}
\end{table}

In Tables \ref{cut_anal1_13}(\ref{cf1}), we display the effect,
  for $\sqrt{s} = 13(14)$~TeV, that these cuts have on the signal and
  background events, when applied successively in the order described
  above. It is noteworthy that a discovery in this final state
  is possible at the current LHC Run with an integrated luminosity as little as 
$\sim 8~{\rm fb^{-1}}$ and $\sim 20~{\rm fb^{-1}}$ for BP1 and BP2
respectively.  The corresponding numbers for LHC Run II
are $\sim 5~{\rm fb^{-1}}$ and $\sim 12~{\rm
  fb^{-1}}$ respectively.


\subsubsection{$1\ell^\pm + 2b + j + \mET $ }
\label{subsec:ii}

Since the interest in this channel owes to the possibility 
of reconstructing the Higgs (and possibly develop an experimental
handle on the very structure of the theory), the cuts now have 
to be reorganized keeping in mind both the origin (and, hence, the 
distributions) of the $b$-jets, as well as the signal strength.
\begin{enumerate}
\item $p_T(\ell_1) > 20$ GeV (C2--1): A single isolated lepton is
  required with a $p_T$ more than 20 GeV.

\item $|\eta(j_1)| < 2.5 ~\&~ p_T(j_1) > 250 $ GeV (C2--2): 
This is exactly akin to cut C1--2 of Sec.~\ref{subsec:i} and owes to the 
fact that the origin of the hardest jet is the same for the two configurations.

\item $\mET > 250$ GeV (C2--3): This, again, is similar to cut C1--5 of 
Sec.~\ref{subsec:i}, and particularly helps eliminate much of the dominant 
$t\bar t$ background.

\item $p_T(b_2) > 40$ GeV (C2--4): As far as the signal events are
  concerned, the $b$-jets arise from the decay chain $Z_h \to A_h + H
  \to A_h + b \bar b$. The large mass difference between the $Z_h$ and
  $A_h$ would be manifested in a large boost for the $H$ which, very
  often, would be translated to a large $p_T$ for the $b$-jets. On the
  other hand, the SM background is dominated by the $t\bar t$ contribution,
  with typically, has a smaller $p_T$ for the $b$-jets. Thus, requiring that 
  {\em at least two} $b$-jets have substantial $p_T$ would discriminate 
  against the background.
    It might seem that imposing a harder cut on $p_T(b_1)$ would be 
    beneficial. While this, {\em per se},  is indeed true, such a gain 
    is subsumed (and, in fact, bettered) by the next two cuts. Hence, we 
    desist from imposing one such.

\item ${\Delta R}_{b \bar b} < 1.5$ (C2--5): {The aforementioned 
large boost for the $H$ in the signal events would, typically, result in the 
two $b$-jets being relatively close to each other. On the other hand, 
the background events from $t\bar t$ would have a much wider distribution, 
whereas $b$'s emanating from associated $H$-production (which, in the SM,
is dominated by low-$p_T$ Higgs) would, preferentially, be back to back (see 
third row of Fig.~\ref{met}).
  Thus, an upper limit on} the angular separation between
  the two tagged $b$-jets  {considerably improves the signal-to background ratio}. 

\item Additional cut $110< M_{b\bar b} < 160$ GeV (C2--6):  This, obviously, serves to 
accentuate the contribution from an on-shell Higgs.	
\end{enumerate}
\def\nl{\cline{1-7}&&&&&&&\\[-2.5ex]}
\begin{table}[ht!]
\centering
\footnotesize
\begin{tabular}{|p{1.7cm}|p{2.1cm}|c|c|c|c|c|c|}
	\hline
	\multicolumn{2}{|c|}{}& \multicolumn{5}{|c|}{Effective Cross-section(fb) after the cut} &${\cal L}_{5\sigma}$ \\ \hline
	SM-background & Production Cross-section (fb) 
	& C2--1 & C2--2 & C2--3 & C2--4 & C2--5 & 
		 \\ \nl
t $+$ jets &  $2.22 \times 10^5$ & \t{3.64}{4} & 685  & 54.7 & 2.7 & 0.56 &  \\ \nl
	t$\bar{\rm t} +$ jets & $7.07 \times 10^5$ & \t{1.93}{5} & \t{1.05}{4} & 826 & 113 & 16.8 &   \\ \nl
	W $+$ jets & $1.54 \times 10^8$ & \t{1.53}{7} & \t{1.73}{4} & \t{1.99}{3} & 0 & 0  &  \\ \nl
	Z $+$ jets & $4.54 \times 10^7$ & \t{2.12}{6} & \t{4.12}{3} & 89.9 & 0 & 0 & \\ \nl
	WW $+$ jets & \t{8.22}{4} & \t{2.21}{4} & 438 & 41.6 & 0.45 & 0 &  \\ \nl
	ZZ $+$ jets & \t{1.10}{4} & \t{1.22}{3} & 27.8 & 1.8 & 0 & 0 &  \\ \nl
	WZ $+$ jets & \t{3.81}{4} & \t{6.73}{3} & 398 & 54.7 & 0.18 & 0.09 &  \\ \nl
	t$\bar{\rm t}+$W & 351 & 131 & 13.5 & 1.8 & 0.16 & 0.02 &  \\ \nl
	t$\bar{\rm t}+$Z & 585 & 189 & 24.6 & 2.8 & 0.3 & 0.06 &  \\ \nl
	t$\bar{\rm t}+$H & 400 & 128 & 14.5 & 1.3 & 0.2 & 0.05 &  \\ \cline{1-7}
	Total background & \multicolumn{5}{|c|}{ } & 17.5 &  \\ \hline \hline
	BP1 & 205 & 64.8 & 58.4 & 46.4 & 2.3 & 1.5 & 211.1 $\rm fb^{-1}$ \\ \hline
	BP2 &  153 & 38.1 & 34.3 & 27.7 & 1.2 & 0.82 & 681.1 $\rm fb^{-1}$ \\  \hline
\end{tabular}
\begin{tabular}{c|p{2.5cm}|c|}
\cline{2-3}
& Effective Cross-section(fb) after additional cut (C2--6) &  ${\cal L}_{5\sigma}$   \\ 
\cline{1-2}	
\multicolumn{1}{|c|}{Total SM background} & 4.8 &  \\ \hline
\multicolumn{1}{|c|}{BP1} & 0.86 & 191.3 $\rm fb^{-1}$ \\ \hline 
\multicolumn{1}{|c|}{BP2} & 0.49 & 551 $\rm fb^{-1}$ \\ \hline
\end{tabular}	
\caption{\it Cut-flow for final state $(ii)$ at $\sqrt{s}= 13$ TeV.}
\label{cut_anal2_13}
\end{table}

\begin{table}[ht!]
\centering
\footnotesize
\begin{tabular}{|p{1.7cm}|p{2.1cm}|c|c|c|c|c|c|}
\hline
\multicolumn{2}{|c|}{}& \multicolumn{5}{|c|}{Effective Cross-section(fb) after the cut} & ${\cal L}_{5\sigma}$ 
\\ \hline
SM-background & Production Cross-sec. (fb) 
	& C2--1 & C2--2 & C2--3 & C2--4 & C2--5 & 
\\ \nl
t $+$ jets & \t{2.49}{5} & \t{4.09}{4} & 769 & 61.4 & 3.0 & 0.63 &   \\ \nl
t$\bar{\rm t} +$ jets & \t{7.96}{5} & \t{2.18}{5} & \t{1.18}{4} & 929 & 127 & 18.9 &   \\ \nl
W $+$ jets & \t{1.66}{8}  & \t{1.64}{7} & \t{1.85}{4} & \t{2.14}{3} & 0  & 0  & \\ \nl
Z $+$ jets & \t{4.86}{7} & \t{2.27}{6} & \t{4.42}{3} & 96.4 & 0 & 0 &   \\ \nl
WW $+$ jets & \t{9.04}{4} & \t{2.43}{4} & 481 & 45.8 & 0.50 & 0 & \\ \nl
ZZ $+$ jets & \t{1.18}{4} & \t{1.31}{3} & 29.7 & 1.9 & 0 & 0 & \\ \nl
WZ $+$ jets & \t{4.21}{4} & \t{7.44}{3} & 439 & 60.5 & 0.20 & 0.12 &   \\ \nl
t$\bar{\rm t}+$W & 398 & 149 & 15.3 & 2.1 & 0.2 & 0.03 &  \\ \nl
t$\bar{\rm t}+$Z & 706 & 228 & 29.8 & 3.4 & 0.4 & 0.07 &  \\ \nl
t$\bar{\rm t}+$H & 479 & 154 & 17.4 & 1.6 & 0.3 & 0.06 &  \\ \cline{1-7}
Total background & \multicolumn{5}{|c|}{ } & 19.7  & \\ \hline \hline
BP1 & 275 & 86.2 & 77.7 & 61.8 & 3.2 & 2.0 & 135.6 $\rm fb^{-1}$ \\ \hline
BP2 &  208 & 51.6 & 46.3 & 37.5 & 1.7 & 1.2 & 362.8  $\rm fb^{-1}$ \\  \hline
\end{tabular}
\begin{tabular}{c|p{2.5cm}|c|}
\cline{2-3}
& Effective Cross-sec. (fb) after additional cut (C2--6) &  ${\cal L}_{5\sigma}$ \\ \cline{1-2}
\multicolumn{1}{|c|}{Total SM background} & 5.4 &  \\ \hline
\multicolumn{1}{|c|}{BP1} & 1.2 & 123.8 $\rm fb^{-1}$ \\ \hline 
\multicolumn{1}{|c|}{BP2} & 0.7 & 319.8 $\rm fb^{-1}$ \\ \hline
\end{tabular}	
\caption{\it Cut-flow for final state $(ii)$ at $\sqrt{s}= 14$ TeV.}
\label{cf2}
\end{table}

The effects of the aforementioned cuts are summarised 
in Tables \ref{cut_anal2_13}(\ref{cf2}). As expected, the signal strength 
is much weaker when compared to that discussed in Sec.~\ref{subsec:i}. While 
the background rate suffers a suppression too, it is not enough and 
the required integrated luminosity is much larger in the present case. 
However, the combination of cuts on $\Delta R_{b \bar b}$ and $M_{b\bar b}$
brings discovery into the realm of possibility even for the present run of 
the LHC and certainly so for that at $\sqrt{s} = 14$~TeV.

\subsubsection{$2\ell^\pm + nj + \mET $ ; $n \geq 2 $}
\label{subsec:iii}
This final state receives contributions only from the 
primary production channels of Eq.~(\ref{s1} \& \ref{s2}), 
and not from that of Eq.~(\ref{s3}). Consequently, the signal 
size is smaller.
 However, the higher multiplicity of the charged lepton
in the final state  
proves helpful in suppressing the background,
provided we re-tune the  kinematic selections as follows:
\begin{enumerate}
\item $|\eta_j| < 2.5$ (C3--1): 
The requirement of jet ``centrality'' remains the same.

\item $p_T(j_1) > 300 ~\rm GeV$ (C3--2): 
The requirement on the hardest jet is now strengthened.
This reduces the cross-sections for most of the 
background subprocesses  by 2-3 order of magnitude whereas the signal 
cross-section is reduced only by a few percent.

\item $p_T(j_2) > 200 ~\rm GeV$ (C3--3):
The preceding cut (C3--2) also serves to harden 
the spectrum of the next sub-leading jet. 
Although this happens for both signal and background, 
the effect is larger for the former. 
This allows us to demand that the next to leading jet 
should have a minimum $p_T$ of 200 GeV. While 
the improvement is not as dramatic as in the case of C3--2, 
the signal to noise ratio does improve fairly (see Table.~\ref{cut_anal3_13}(\ref{cf3})).

\item $p_T(\ell_1) > 20 ~\rm GeV$ (C3--4): 
As in the previous cases, this requirement on
the hardest isolated lepton 
is a very good discriminant. Most of the major background 
sources are suppressed by at least an order of magnitude (see Table.~\ref{cut_anal3_13}(\ref{cf3})),
with the suppressions for the single-top and $Z + n$-jets production 
being even more pronounced.

\item $p_T(\ell_2) > 20 ~\rm GeV$ (C3--5):  
The success of C3--4 prompts us to require that even the 
next hardest isolated lepton should have $p_T > 20$ GeV. As Table.~\ref{cut_anal3_13}(\ref{cf3})
testifies, the major backgrounds (after the imposition of C3--4), namely
 $t\bar t$-- and $W + n$-jets production suffer severe suppression.

\item $\rm M_{eff} > 1.2$ TeV (C3--6): 
Analogous to (C1--4), we impose a high value of $\rm M_{eff}$ to 
further	suppress the background cross section.

\item $\mET > 250$ GeV (C3--7): Additionally, we demand a large $\mET$
to enhance the signal significance.
\end{enumerate} 
\def\nl{\cline{1-9}&&&&&&&&&\\[-2.5ex]}
\begin{table}[ht!]
\centering
\footnotesize
\begin{tabular}{|p{1.7cm}|p{2.1cm}|c|c|c|c|c|c|c|c|}
\hline
\multicolumn{2}{|c|}{}& \multicolumn{7}{|c|}{Effective Cross-section(fb) after the cut} & ${\cal L}_{5\sigma}$ \\ \hline
SM-background & Production Cross-sec. (fb) & C3--1 & C3--2 & C3--3 & C3--4 & C3--5 & C3--6 & C3--7 &   \\ \nl
t $+$ jets & \t{2.22}{5} & \t{1.43}{5} & \t{2.54}{3} & \t{1.75}{3} & 104 & 0 & 0 & 0 & \\ \nl
t$\bar{\rm t} +$ jets & \t{7.07}{5} & \t{6.09}{5} & \t{3.21}{4} & \t{2.11}{4} & \t{2.82}{3} & 99 & 22.8 & 4.2 & \\ \nl
W $+$ jets & \t{1.54}{8} & \t{4.43}{7} & \t{7.47}{4} & \t{5.47}{4} & \t{3.48}{3} & 0  & 0 & 0 & \\ \nl
Z $+$ jets & \t{4.54}{7} & \t{1.31}{7} & \t{3.52}{4} & \t{2.73}{4} & 737 & 270 & 18.0 & 0 & \\ \nl
WW $+$ jets & \t{8.22}{4} & \t{6.44}{4} & \t{1.13}{3} & 751 & 59 & 1.3 & 0.11 & 0.11 & \\ \nl
ZZ $+$ jets & \t{1.10}{4} & \t{6.31}{3} & 154 & 111 & 6.5 & 2.9 & 0.60 & 0.04 & \\ \nl
WZ $+$ jets & \t{3.81}{4} & \t{2.21}{4} & \t{1.28}{3} & 907 & 101 & 14.4  & 3.2 & 0.55 & \\ \nl
t$\bar{\rm t}+$W & 351 & 308 & 35.4 & 22.3 & 4.5 & 0.1 & 0.05 & 0.03 &  \\ \nl
t$\bar{\rm t}+$Z & 585 & 532 & 76.9 & 48.4 & 8.3 & 0.9 & 0.17 & 0.06 &  \\ \nl
t$\bar{\rm t}+$H & 400 & 369 & 50.1 & 33.1 & 6.0 & 0.4 & 0.18 & 0.08 &  \\ \cline{1-9}
Total background & \multicolumn{7}{|c|}{ } & 5.1 &  \\ \hline \hline
BP1 & 205 & 198 & 186 & 163 & 42.8 & 3.1 & 2.4 & 2.1 & 40.8 $\rm fb^{-1}$ \\ \hline
BP2 &  153 & 149 & 141 & 126 & 25 & 1.3 & 0.96 & 0.84 & 210.4 $\rm fb^{-1}$ \\  \hline
\end{tabular}
\caption{\it Cut-flow for final state $(iii)$ at $\sqrt{s}= 13$ TeV.}
\label{cut_anal3_13}
\end{table}

\begin{table}[ht!]
\centering
\footnotesize
\begin{tabular}{|p{1.7cm}|p{2.1cm}|c|c|c|c|c|c|c|c|}
\hline
\multicolumn{2}{|c|}{}& \multicolumn{7}{|c|}{Effective Cross-section(fb) after the cut} & ${\cal L}_{5\sigma}$\\ \hline
SM-background & Production Cross-sec. (fb) & C3--1 & C3--2 & C3--3 & C3--4 & C3--5 & C3--6 & C3--7 & 
  \\ \nl
t $+$ jets & \t{2.49}{5} & \t{1.60}{5} & \t{2.85}{3} & \t{1.96}{3} & 116 & 0 & 0 & 0 & \\ \nl
t$\bar{\rm t} +$ jets & \t{7.96}{5} & \t{6.86}{5} & \t{3.62}{4} & \t{2.37}{4} & \t{3.17}{3} & 111 & 25.7 & 4.7 & \\ \nl
W $+$ jets & \t{1.66}{8} & \t{4.75}{7} & \t{8.00}{4} & \t{5.86}{4} & \t{3.72}{3} & 0  & 0 & 0 & \\ \nl
Z $+$ jets & \t{4.86}{7} & \t{1.40}{7} & \t{3.77}{4} & \t{2.93}{4} & 790 & 289 & 19.3 & 0 &\\ \nl
WW $+$ jets & \t{9.04}{4} & \t{7.08}{4} & \t{1.24}{3} & 826 & 64.9 & 1.4 & 0.13 & 0.13 &\\ \nl
ZZ $+$ jets & \t{1.18}{4} & \t{6.74}{3} & 164 & 118 & 6.9 & 3.1 & 0.64 & 0.04 &\\ \nl
WZ $+$ jets & \t{4.21}{4} & \t{2.45}{4} & \t{1.41}{3} & \t{1.02}{3} & 112 & 15.9 & 3.5 & 0.61 &  \\ \nl
t$\bar{\rm t}+$W & 398 & 349 & 40.1 & 25.3 & 5.1 & 0.2 & 0.06 & 0.04 &  \\ \nl
t$\bar{\rm t}+$Z & 706 & 642 & 92.9 & 58.4 & 10.0 & 1.1 & 0.21 & 0.07 &  \\ \nl
t$\bar{\rm t}+$H & 479 & 443 & 60 & 39.7 & 7.2 & 0.5 & 0.22 & 0.1 &  \\ \cline{1-9}
Total background & \multicolumn{7}{|c|}{ } & 5.7 &  \\ \hline \hline
BP1 & 275 & 266 & 249 & 219 & 57.2 & 4.3 & 3.2 & 2.8 & 27.1 $\rm fb^{-1}$ \\ \hline
BP2 &  208 & 201 & 190 & 171 & 34.5 & 1.9 & 1.4 & 1.2 & 119.8 $\rm fb^{-1}$ \\  \hline
\end{tabular}
\caption{\it Cut-flow for final state $(iii)$ at $\sqrt{s}= 14$ TeV.}
\label{cf3}
\end{table}

Note that although the dilepton final state has a reduced signal 
cross-section as compared to that with a single lepton, the requirement
of a second isolated lepton also significantly reduces the background. 
Therefore, this final state requires moderate values
for integrated luminosity at 13(14) TeV LHC, namely around
$\sim 40(26)$ and $\sim 203(116)~ \rm fb^{-1}$ for BP1 and BP2 respectively,
which would be accessible in the current run of LHC.

At this point, we would like to mention that the benchmark points
considered in our analysis can also be probed via the pair
production of heavy $T$-odd gauge bosons ($W_H/Z_H$). However,
the later processes being purely electroweak in nature yields
much lower cross-section which in turn require significantly
higher luminosity to reach the same signal significance as ours.
This has been studied in Ref.~\cite{Cao:2015cdb}.

\section{Summary and conclusions}\label{conclusion}

The very lightness of the Higgs boson that was discovered at the LHC
has been a cause for concern, especially in the absence of any
indication for physics beyond the SM that could be responsible for
keeping it light.  Amongst others, Little Higgs scenarios provide an
intriguing explanation for the same.  While several variants have been
considered in the literature, in this paper, we examine a particularly
elegant version, namely the Littlest Higgs model with a $Z_2$ symmetry
($T$-parity). The latter not only alleviates the severe constraints
(on such models) from the electroweak precision measurements but also
provides for a viable Dark Matter candidate in the shape of $A_h$, the
exotic gauge partner of the photon.

At the LHC, the exotic particles can only be pair-produced on account
of the aforementioned $T$-parity. Understandably, the production cross
sections are, typically, the largest for the strongly-interacting
particles. For example, if the exotic quarks $Q_{ih}$ are light enough
to have a large branching fraction into their SM counterparts and the
$A_h$, we would have a very pronounced excess in a final state
comprising a dijet along with large
missing-$p_T$~\cite{Choudhury:2012xc}. On the other hand, if the
$Q_{ih}$ are heavier than $W_h$ and $Z_h$ (as can happen for a wide
expanse of the parameter space) then they decay into the latter
instead, with these, in turn decaying into their SM counterparts (or
the Higgs), resulting in a final state comprising multiple jets,
possibly leptons and missing-$p_T$~\cite{Choudhury:2006mp,Cacciapaglia:2009cu}, and it is
this possibility that we concentrate on.  
The parameter space of interest is the two-dimensional one, 
spanned by $f$, the scale of
breaking of the larger symmetry and $\kappa$, the universal Yukawa
coupling. Although a part of it is already ruled out by the 
negative results from the 8 TeV run, a very large expanse 
is still unconstrained by these analyses. We illustrate 
our search strategies choosing representative benchmark points from within
the latter set. We consider not
only the production of a pair of exotic quarks, but also the
associated production of $W_h/Z_h$ with such a quark.  Concentrating
on the final state comprising leptons plus jets plus missing
transverse energy, we consider all the SM processes that could
conspire to contribute as background to our LHT signal, and perform a
full detector level simulation of the signal and background to
estimate the discovery potential at the current run and subsequent
upgrade of the LHC.

The large mass difference between the $Q_{ih}$ and $W_h/Z_h$ results
in large momenta for at least a few of the jets. Similarly, the even
larger mass difference between the $W_h/Z_h$ and the $A_h$ results,
typically, in large missing-momentum.  This encourages us to consider
final states consisting of hard jets and leptons and large missing
transverse momentum.  We observe that final states with only one
isolated charged lepton ($e^\pm, \mu^\pm$) and with at least three
jets and substantial missing transverse energy are the ones most
amenable to discovery.  For example, at the 13(14) TeV LHC with only
$8(5)~\rm fb^{-1}$ and $20(12)~\rm
fb^{-1}$ integrated luminosity for our BP1 and BP2 respectively.  A
{\em confirmatory test} is afforded by a final state requiring one
extra isolated lepton. Though this decreases the signal cross-section
significantly, LHC Run II can still reach the discovery level, but now
only with $40(26)~ \rm fb^{-1}$ and $203(116)~ \rm
fb^{-1}$ integrated luminosity at 13(14) TeV.

We however wish to highlight, through this work, a more interesting
signal, one with 2 tagged $b$-jets in the final state.  As discussed
earlier (Eq.~(\ref{dk})), the heavy $Z_h$ boson decays to a Higgs boson
and the $A_h$ with almost 100\% branching ratio.  This presents us
with an unique opportunity to reconstruct the Higgs mass from the
tagged b-jets, thus providing us with an important insight into the
LHT parameter space. As we have found out, the reconstruction of Higgs
mass requires higher integrated luminosity (few hundred $\rm fb^{-1})$)
but it is still within the reach of the LHC run II.  We, thus, hope
that our analysis demonstrates the viability of testing the LHT model
in the current run of the LHC.

\medskip
\emph {Acknowledgments:} 
The work of S.K.R. was partially supported by funding available from
the Department of Atomic Energy, Government of India, for the Regional
Centre for Accelerator-based Particle Physics (RECAPP), Harish-Chandra
Research Institute. D.C. acknowledges partial support from the
European Union FP7 ITN INVISIBLES (Marie Curie Actions,
PITN--GA--2011--289442), and the Research and Development grant of the
University of Delhi. D.K.G and I.S would like to acknowledge the hospitality
of RECAPP (HRI) in various occasions during the work. I.S would like to thank
Juhi Dutta, Sujoy Poddar and Subhadeep Mondal for useful 
discussions and technical help. 
Computational work for this study was partially carried out at the cluster
computing facility in the RECAPP.

\bibliographystyle{JHEP}
\bibliography{ref.bib}
\end{document}